\documentstyle[aps]{revtex}
\begin{document}
\title{Identification and collapse---the seeds of quantum superluminal communication}
\author{Gao Shan}
\address{Institute of Quantum Mechanics, 11-10, NO.10 Building,
YueTan XiJie DongLi, XiCheng District, Beijing 100045, P.R.China.
Email: gaoshan.iqm@263.net}
\date{June 14, 1999}
\maketitle

\vskip 0.5cm

\begin{abstract}

We deeply analyze the relation between the identification of the
classical definite states and collapse of the superposition state
of such definite states during identification, and show that
although identification rejects superposition, it needs not result
in collapse, which then provides one possibility to achieve
quantum superluminal communication.

\end{abstract}

\vskip 0.2cm

The special role of measurement was first stressed by
Bohr\cite{Bohr} in his complementarity principle to consistently
interpret the non-classicality of the quantum world, then von
Neumann\cite{Neumann} rigorously formulated the measurement
process in his measurement theory involving projection postulate,
but the inherent fuzziness in their definition of measurement or
projection still exists, thus in order to end the infinite
spreading of linear superposition, identification of the observer
is implicitly resorted by von Neumann\cite{Neumann} and further
advocated by Wigner\cite{Wigner} to break the linear superposition
and generate the definite result, this may be the first statement
about the relation between identification and collapse, it is
simply if identified then collapse.

But, when facing the problem of quantum cosmology, this relation
needs to be greatly revised, since for the state of the whole
universe, no outside measurement device or observer exists, the
special role of measurement and identification are essentially
deprived, and the collapse process, if exists, must be the own
thing of the wave function. The recent dynamical collapse
theories\cite{Ghir86}\cite{Pea86}\cite{Pen86}\cite{Dio87}
\cite{Dio89}\cite{Pea89}\cite{Ghir90}\cite{Pen96}\cite{Gao99}further
revise the above relation, according to which the normal linear
evolution and projection process of the wave function are unified
in one revised stochastic Schr\"{o}dinger equation, the collapse
process is just the natural result of such evolution, thus the new
relation is whether or not identified collapse will happen.

Although collapse needs not resort to the identification of the
observer, and essentially an objective process, people still
implicitly stick to the orthodox view, which asserts that after
the conscious observer can identify the classical definite state,
the collapse of the observed superposition state of such states
must happen, and have still tried to demonstrate that according to
the dynamical collapse theory, our brain just satisfies this
condition\cite{Penrose1}, thus it appears that identification is
essentially connected with collapse again, which is evidently not
accounted for by the dynamical collapse theory itself, and de
facto results only from the requirement of the orthodox
interpretation of present quantum theory.

On the other hand, the wide acceptance of this orthodox view is
also due to its confusion with the well-known conclusion that
identification rejects superposition, which states that when the
conscious observer can identify the measurement result of the
superposition state, then collapse must happen, where the
identification time denotes the time to identify the definite
result for the measured superposition state, not for the measured
classical definite state, while this conclusion is reasonable,
since it only means that, on the one hand, if the observed
superposition state is identified as a definite result or
perception, namely the identification part of the conscious
observer is in a definite state, then collapse must happen; on the
other hand, if the identification part of the conscious observer
is in a superposition state, then no definite perception exists,
and no definite result is identified either. ( It is difficult to
accept that when identified as a definite result the superposition
state still exists. )

In the following, we will physically re-analyze the relation
between the identification and collapse on the basis of the
dynamical collapse theory, and show that although identification
rejects superposition, it needs not result in collapse, and their
combination can result in quantum superluminal
communication\cite{Gao1}. First, if the collapse process of the
measured superposition state is completed before the conscious
observer identifies the measurement result, which may result from
the entanglement of the measuring device, then the identification
is indeed irrelevant to the collapse process, since what he
identifies is just a classical definite state; Secondly, if the
collapse process of the measured superposition state is not
completed before the conscious observer identifies the measurement
result, then the identification process of the conscious observer
will surely influence the collapse process of the measured
superposition state, especially in an extreme situation, if the
conscious observer is the only "measuring device", then the
collapse process will be mainly determined by the identification
process, we will further analyze this situation in detail.

On the one hand, as to the identification process of a conscious
observer about a classical definite state, which is one of the
states in the above measured superposition state, there are mainly
two physical properties characterizing the process, one is the
entangled energy to identify the state, the other is the
identification time after which the result is identified, and in
general there exists no essential relation between them, but it is
reasonable that with the natural selection, in which only the
classical definite states are input to the conscious observer, the
entangled energy will turn to be smaller and smaller, and the
identification time will turn shorter and shorter.

On the other hand, according to the general dynamical collapse
theories, if the entangled energy turns to be small, then the
collapse time will turn to be long, then it is reasonable that
with the natural evolution there must appear the conscious
observer, for which the collapse of the observed superposition
state indeed happens after the relevant classical definite state
is identified, and his collapse time, or identification time, for
identifying some superposition state is much longer than his
identification time for identifying one of the definite classical
states in the corresponding superposition state, so that such
conscious observer can be conscious of the time difference of
these two identifications, and distinguish the measured
non-orthogonal single states, then it will be an easy thing to
achieve quantum superluminal communication\cite{Gao1}.

On the whole, we have shown that although identification rejects
superposition, it needs not result in collapse, and this just
provides one possibility to achieve quantum superluminal
communication.

\end{document}